\documentclass[aps,pra,twocolumn,superscriptaddress,nofootinbib]{revtex4-2}

\usepackage[english]{babel}
\usepackage[utf8]{inputenc}
\usepackage[T1]{fontenc}

\usepackage{amsmath,amssymb}
\usepackage{graphicx}
\usepackage{tikz}   
\usetikzlibrary{quantikz2,positioning}
\usepackage{longtable}
\usepackage{booktabs}
\usepackage{adjustbox}
\usepackage{subfigure}
\usepackage{tabularx}
\usepackage[colorlinks=true, allcolors=blue]{hyperref}

\begin{document}

\title{Efficient Quantum Information-Inspired Ansatz for Variational Quantum Eigensolver Algorithm: Applications to Atomic Systems} 

\author{Abdul Kalam}
\email{ak6994231@gmail.com}
\affiliation{Centre for Quantum Engineering, Research and Education, TCG Crest, Kolkata 700091, India}
\affiliation{Academy of Scientific and Innovative Research (AcSIR), Ghaziabad 201002, India}

\author{Prasenjit Deb}
\email{devprasen@gmail.com}
\affiliation{Centre for Quantum Engineering, Research and Education, TCG Crest, Kolkata 700091, India}

\author{Akitada Sakurai}
\email{akitada.sakurai@oist.jp}
\affiliation{Okinawa Institute of Science and Technology Graduate University, Onna-son, Okinawa 904-0495, Japan}

\author{B. K. Sahoo}
\email{bijayakumar.sahoo@gmail.com}
\affiliation{Atomic, Molecular and Optical Physics Division, Physical Research Laboratory, Navrangpura, Ahmedabad 380009, India}

\author{V. S. Prasannaa}
\email{srinivasaprasannaa@gmail.com}
\affiliation{Centre for Quantum Engineering, Research and Education, TCG Crest, Kolkata 700091, India}
\affiliation{Academy of Scientific and Innovative Research (AcSIR), Ghaziabad 201002, India}

\author{B. P. Das}
\email{Bhanu.das@tcgcrest.org}
\affiliation{Centre for Quantum Engineering, Research and Education, TCG Crest, Kolkata 700091, India}
\affiliation{Academy of Scientific and Innovative Research (AcSIR), Ghaziabad 201002, India}
\affiliation{Department of Physics, Institute of Science Tokyo, 2-12-1 Ookayama, Meguro City, Tokyo 152-8550, Japan}


\begin{abstract} 
We present a quantum information-inspired ansatz for the variational quantum eigensolver (VQE) and demonstrate its efficacy in calculating ground-state energies of atomic systems. Instead of adopting a heuristic approach, we start with an approximate multi-qubit target state and utilize two quantum information-theoretic quantities, i.e., von Neumann entropy and quantum mutual information, to construct our ansatz. The quantum information encoded in the target state helps us to design unique blocks and identify qubit pairs that share maximum quantum correlations among them in the multi-qubit system, thereby enabling us to deterministically place two-qubit entanglers in the suitably constructed parametrized quantum circuit. We find that our approach has the advantage of reduced circuit depth compared to the unitary coupled-cluster (UCC) ansatz (the gold standard for VQE), and yet yields accurate results. To test the performance of our ansatz, we apply it to compute ground-state energies of atomic systems.
We find that for up to 12 qubits (or 12 spin orbitals) noiseless calculation, the proposed ansatz yields energies with \(99.99\%\) accuracy relative to the complete active space configuration interaction values, while utilizing only two blocks, which contain at most \(99 \%\) fewer 2-qubit gates than the UCC ansatz.
\end{abstract}

\maketitle

Quantum computation promises a wide range of applications in both basic science and real-world problems. One of the important motivations underlying the development of quantum computers and algorithms is the simulation of quantum phenomena in atoms, molecules, and condensed-matter systems, to name a few \cite{aspuru2005simulated,peruzzo2014variational,cao2019quantum,mcardle2020quantum,baskaran2023adapting,cade2020strategies,shen2017quantum,hempel2018quantum,barkoutsos2018quantum}, and studying various physical and chemical properties of these systems. An example of this is the computation of ground-state energies of atoms and molecules, which are of particular interest in quantum physics and chemistry. In recent years, considerable efforts have gone into developing and applying quantum algorithms to investigate this property. Among these algorithms, the ones based on VQE have been the most widely used \cite{guo2024experimental,zhao2023orbital,gao2021applications,tilly2022variational,chawla2025relativistic,singh2024experimental,guo2024experimental,chawla2025relativistic, tang2021qubit,yordanov2021qubit,chawla2025vqe}. VQE exploits both classical and quantum resources, making it a hybrid quantum-classical algorithm. It has attracted much attention due to the feasibility of its implementation with a relatively few qubits on Noisy Intermediate-Scale Quantum (NISQ) devices \cite{ying2023experimental, kandala2017hardware, kandala2019error}. Fault-tolerant era quantum algorithms \cite{abrams1997simulation,abrams1999quantum}, aimed for the same purpose, lack the advantages provided by VQE, making them unsuitable for NISQ-era quantum computing. 
The starting point of VQE is the preparation of a trial wavefunction, which is based on an \textit{ansatz} and is represented by the action of a parameter-dependent unitary operator on a reference state. This wavefunction is then used to evaluate the energy functional on a quantum computer, while the parameter update is performed on a classical computer.
Broadly, there are two choices for the VQE ansatz; one of them is based on the unitary coupled-cluster (UCC) theory \cite{romero2018strategies,whitfield2011simulation,evangelista2011alternative,cooper2010benchmark} -- the UCC ansatz (adapted from the coupled-cluster (CC) method \cite{bartlett2007coupled}), and the other one is --the Hardware-efficient ansatz (HEA) \cite{kandala2017hardware,leone2024practical,ratini2022wave,tang2021qubit,rattew2019domain}. The UCC ansatz is physically motivated, but it quickly becomes impractical for large systems on current NISQ hardware due to its polynomial scaling in circuit depth. For instance, the UCC singles and doubles (UCCSD) ansatz exhibits a circuit depth scaling of $O(N^5)$, where $N$ is the number of spin orbitals or qubits \cite{tilly2022variational}. In contrast, HEA consists of layers of single-qubit rotation gates to create superposition and two-qubit gates to create entanglement, without any information of the chemical system \cite{sim2019expressibility}. HEAs, though shallow and compatible with current quantum hardware, suffer from limitations due to heuristic nature of their constructions. Firstly, shallow circuits may fail to cover the relevant Hilbert space sufficiently, while deeper circuits with a larger number of qubits could encounter the barren plateau problem \cite{mcclean2018barren,holmes2022connecting}. Secondly, the fixed pattern of entanglement in HEAs may not be universally suitable, leading to either lower or excessive entanglement in the system of interest \cite{sim2019expressibility}. 
In addition to the circuit depth of the ansatz, the number of measurements required on NISQ hardware also impacts performance. Since the atomic or molecular Hamiltonian \( H \) typically contains \(\mathcal{O}(N^4)\) Pauli terms, an equivalent number of measurements may be needed. To reduce this overhead, several strategies have been explored, including qubit pair-wise commutation \cite{cadi2024folded} and employing joint Bell-basis measurements \cite{cao2024accelerated}.


In this work, leveraging two quantum information-theoretic quantities -- \textit{von Neumann entropy} ($S$) \cite{quantum_info_wilde} and \textit{quantum mutual information} ($I$) \cite{quantum_info_wilde}, we construct a suitable HEA ansatz for VQE. Unlike traditional HEAs, which often rely on arbitrary arrangement of entangling gates between two qubits, our quantum information-inspired HEA (hereafter QIIA), inspired by the earlier work from references \cite{tkachenko2021correlation,qida,materia2024quantum,zhang2021mutual}, is systematically constructed based on the quantum information-theoretic quantities mentioned above and solves the arbitrary arrangement of entangling gates in the ansatz. To check the efficiency of our proposed ansatz, we apply it to relativistic atomic physics and evaluate ground-state energies of different atoms. We find that with only two blocks, the QIIA ansatz achieves 99.99\% accuracy in the noiseless medium relative to the standard complete active space configuration interaction (CAS-CI) value. As a proof of principle, we executed the QIIA circuit on the IBM Sherbrooke quantum hardware, incorporating resource reduction techniques on both the Hamiltonian and the circuit.

\begin{figure*}[tbh]
  \centering
 \includegraphics[scale=.22]{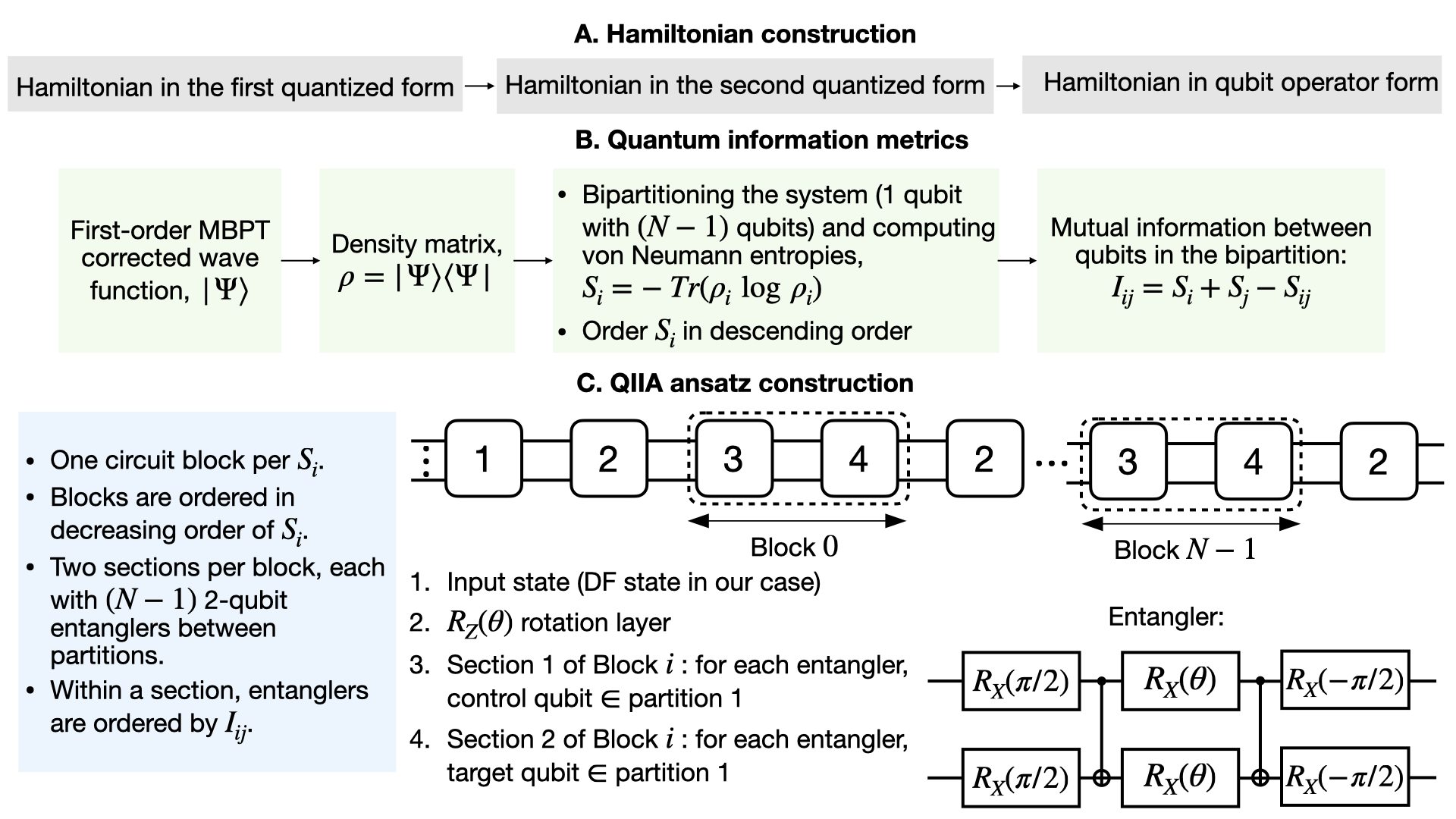}
\caption{
Flowchart illustrating construction of the qubit Hamiltonian and QIIA ansatz, which will be fed into the standard VQE procedure. The relativistic DC Hamiltonian is mapped to a qubit Hamiltonian, and the MBPT first-order corrected wavefunction is used to compute the von Neumann entropy and mutual information between the qubit pairs. These quantum information metrics guide the placement of blocks and the entanglers inside the block. The two-qubit match gate type entangler used in QIIA is also shown in the flowchart.
}
\label{fig:flowchart1}
\end{figure*}

Albeit the QIIA ansatz is applied to the atomic Hamiltonian in this work, it can be easily extended to the molecular Hamiltonian, as well as calculating other properties of both atoms and molecules. 
While the VQE algorithm has been widely applied to molecular systems, its application to atomic systems, particularly in the relativistic framework, remains comparatively underexplored. We therefore begin with the relativistic Dirac-Coulomb (DC) Hamiltonian for atomic systems (see Chapter 6 of Ref.~\cite{grant2007relativistic}): 


\begin{equation}
\begin{aligned}
H_{DC} &= \sum_{i}^{N_{e}} \!\left[ c\,\vec{\alpha}_i \cdot \vec{p}_i + \beta_i c^2 + V_{nuc}(\vec{r}_i) \right] \\
&\quad + \frac{1}{2} \sum_{i \neq j} \frac{1}{|\vec{r}_i - \vec{r}_j|},
\end{aligned}
\label{eq:DC hamiltonian}
\end{equation}

where \( c \) is the speed of light, \( \vec{{p}_i} \) is the momentum of the $i^{th}$ electron, \({N_{e}}\) is the number of electrons and \(\vec{r}_i \) is the coordinate vector of the $i^{th}$ electron. The quantities \( {\alpha}_i \) and \( \beta_i \) are the Dirac spinors, $ V_{nuc}({\vec{r}_i})$ is the electron–nucleus potential of the $i^{th}$ electron (for this work, we considered the Fermi nuclear model). We have used the atomic units in Eq.(\ref{eq:DC hamiltonian}). The initial ground-state is obtained using the Dirac-Hartree-Fock (DF) method (see chapter 6 of Ref. \cite{grant2007relativistic}). By decomposing the Hamiltonian as \( H_{DC} = H_0 + H' \), where \( H_0 \) is the one-body DF Hamiltonian and \( H' \) is the residual Hamiltonian (the difference between the exact two-body coulumb and the DF Hamiltonians), the DF energy is given by \( E_{\text{DF}} = \langle \Phi_0 | H_{DC} | \Phi_0 \rangle \) where \(|\Phi_0 \rangle\) is the DF state. The total energy includes the correlation correction to the \(E_{\text{DF}}\) i.e, \( E_0 = E_{\text{DF}} + E_{\text{corr}} \). 
The second-quantized form of $H_{DC}$ is given by
$H = \sum_{pq} h_{pq} a_p^\dagger a_q + \frac{1}{2} \sum_{pqrs} g_{pqrs} a_p^\dagger a_q^\dagger a_s a_r$,
where \( a_p^\dagger \) for example is a fermionic creation operator where as \( a_q \) is a fermionic annihilation operator. \( h_{pq} \) and \( g_{pqrs} \) are one- and two-electron integrals respectively (page 228 - 230 of Ref. \cite{lindgren2012atomic}). To perform VQE, mapping to a qubit Hamiltonian yields \( H_q = \sum_k w_k P_k \), where \( w_k \) are real coefficients and \( P_k \in \{I, X, Y, Z\}^{\otimes N} \) are Pauli strings. We use the Jordan-Wigner transformation~\cite{seeley2012bravyi} for this mapping. Figure \ref{fig:flowchart1} schematically outlines the Hamiltonian construction and QIIA circuit preparation steps used for the VQE algorithm.

Initial trial states are constructed using many-body perturbation theory (MBPT) with first-order correction to the wavefunction, i.e, second-order correction to the energy (Chapter 9 of Ref.~\cite{lindgren2012atomic}). This theory is capable of capturing electron correlation efficiently for many closed-shell atoms. 
The first-order wavefunction correction \( |\Phi^{(1)}\rangle \) accounts for electron correlation and is given as 
\begin{equation}
|\Phi^{(1)}\rangle = \sum_{I\neq0} \frac{\langle \Phi_I| H' | \Phi_0 \rangle}{E_0 - E_I} |\Phi_I\rangle,
\end{equation}
where \( E_0 \) and \( E_I \) are the DF and excited determinant \(|\Phi_I\rangle \) energies, respectively. The resulting MBPT first-order corrected wavefunction is normalized as \( |\Psi\rangle = \frac{1}{\sqrt{\mathcal{N}}} (|\Phi_0\rangle + |\Phi^{(1)}\rangle) \). 
\(|\Phi_0 \rangle\) and \(|\Phi^{(1)}\) are built from
single-particle orbitals characterized by the quantum numbers 
$(n, l, s, j, m_j)$, where $n$ denotes the principal quantum number, 
$l$ the orbital angular momentum quantum number, 
$s$ the spin quantum number, $j$ the total angular momentum quantum number, 
and $m_j$ is the magnetic quantum number.
The qubits in the state are arranged as a concatenation of two bitstrings, \(\ket{-m_j, +m_j}\). 
As an example, the above-mentioned normalized wavefunction can be expressed in the computational basis state for $B^+$ ion  (2e, 8q) system (\(1s\) orbital is frozen, the \(2s\) orbital is occupied, and the \(2p\) orbital is unoccupied) as:
\begin{align}
\ket{\Psi} &= 
0.996389\,\ket{00010001} 
- 0.020173\, \ket{00100010} \notag \\
&\quad 
- 0.024871\, \ket{01000100} 
+ 0.078577\, \ket{10001000}.
\label{state_in_computational_basis}
\end{align}
The qubit strings representing the basis for these states follow little-endian convention, with the $0^{th}$ qubit ($q_0$) corresponding to the least significant bit. 
We then utilize the MBPT-corrected wavefunctions to study entanglement properties of the systems of interest. Specifically, we first map the many-body system to a set of all possible two-body systems by considering the bipartitions $i^{th}$ qubit versus the remaining qubits. 
Then we compute the von Neumann entropy of every bipartition to identify the highly entangled ones.

 For pure multi-qubit states, the von Neumann entropy of the subsystems is a measure of entanglement and hence, quantum correlations. The von Neumann entropy of a quantum state \(\rho\) is defined as
\begin{equation}
S(\rho) = -\mathrm{Tr}(\rho\log_2 \rho),
\end{equation}
where "$\mbox{Tr}$" represents the trace operation of a matrix. After identifying the bipartitions that have a considerable amount of quantum correlations in the form of entanglement, we calculate quantum mutual information between the $i^{th}$ qubit and each of the remaining qubits. Quantum mutual information is a physical quantity that signifies the total correlation between the subsystems of any correlated quantum state. Being widely used in the quantum information and communication theory, mutual information between the $i^{th}$ and $j^{th}$ qubit can be expressed as    
\begin{equation}
I_{ij} = S_i + S_j - S_{ij} ,
\end{equation}
where  \(S_i\) and \(S_j\) are the von Neumann entropies of  $i^{th} ~\text{and}~ j^{th}$ qubits, respectively, and  \(S_{ij}\) is the von Neumann entropy of the joint state of the qubits. This analysis enables us to construct an entanglement map that reflects the electronic correlation structure in the approximate ground-state considered initially.

\begin{figure}[tbh]
\centering
\subfigure[]{\includegraphics[width=0.238\textwidth]{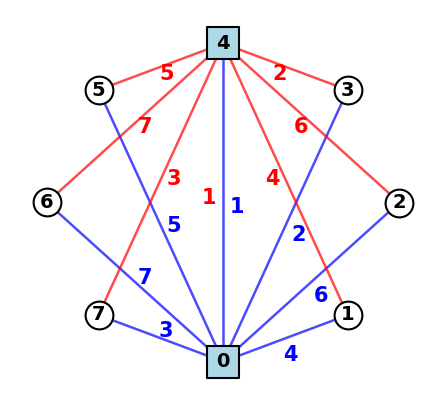}}
\subfigure[]{\includegraphics[width=0.238\textwidth]{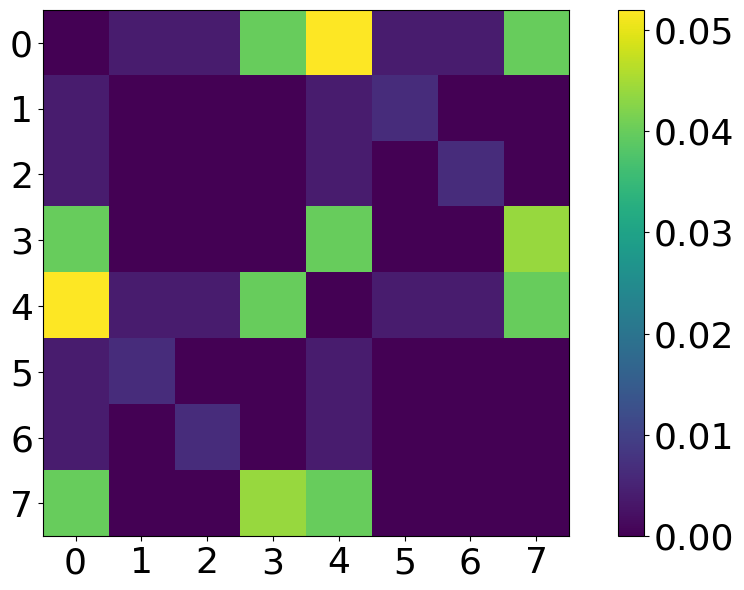}}
\caption{Entanglement structure used for constructing our proposed ansatz.
Sub-figure (a) Graph representation of the QIIA circuit, where nodes denote qubits and edges indicate entangling gates. Only the two bipartitions with the largest von Neumann entropies are selected, yielding two circuit blocks in the ansatz.  The nodes marked inside a square represent those whose entropies are the largest, with the rest of them marked in a circle. The edge weights on the blue coloured edges indicate the order of occurrence of the entanglers in the 0$^{th}$ block, whereas those for the red coloured ones denote the order of occurrence in the 1$^{st}$ block. The order of occurrence is decided by mutual information, as sub-figure (b) shows. The mutual information between two qubits (shown along the x- and y-axes) is colour-coded in the sub-figure. }
\label{fig:figure with graph state}
\end{figure}

Once we capture the information on quantum correlations in a coarse-grained manner from the approximated density, $\rho$, we move on to construct the parametrized circuit. We introduce a variational matchgate-style entangler \cite{bassman2022constant} $U_{ent}(\theta)$ (see Figure \ref{fig:flowchart1}) that preserves the particle number using only two CNOT gates per entangler. More information about the entangler is given in Section 3 of the Supplementary Material. 
For an eight-qubit system consisting of two electrons and eight spin orbitals, Figure \ref{fig:figure with graph state} shows the two highest von Neumann entropies and the mutual information between qubit pairs, illustrating the entangler arrangement in the ansatz through a connectivity graph and a heat map. This information is used to construct the QIIA circuit shown in Figure~\ref{fig:proposed_quantum_circuit} of the Supplementary Material. The circuit begins with the DF state, followed by a layer of $R_z$ rotation. This is succeeded by a structured entangling layer inside multiple blocks. Each block represents a unique bipartition, and the blocks are sorted in descending order from left to right depending on the values of the von Neumann entropy evaluated for the bipartitions (in Figure \ref{fig:proposed_quantum_circuit}, we have only considered one block corresponding to $q_{0}$ ). 
Each block consists of two sections: in the first, the $i^{th}$ qubit acts as the control while the remaining qubits are targets, and the second is just \textit{vice versa}. 
Now, if we take the random permutation approach to place the entanglers in the correct position, the number of possible unique circuits will increase rapidly with system size. For example, if we have taken \(\mathcal{L}\) blocks and each block is arranged in \(\mathcal{P}\) possible ways, then the total number of circuit permutations will be \(\mathcal{P}^\mathcal{L}\). Instead, 
to determine the optimal configuration of the entangling gates between the qubits within a block, we utilize the mutual information data.
This ensures insertion of entangling gates between the highly correlated qubit pairs first, followed by the other pairs in the descending order of mutual information value. Thus, our proposed ansatz not only captures most of the quantum correlations present in the many-body system, but it also excludes the qubit pairs having a negligible amount of entanglement between them, thereby significantly reducing the number of two-qubit gates. All initial wavefunctions used for computing these metrics and how these metrics help to capture the important quantum correlation are provided in Sections 1 and 2 of the Supplementary Material. Notably, the same entangler ordering was obtained using the more accurate and rigorous linearized coupled-cluster method, consistent with that derived from the first-order MBPT-corrected wavefunction.


Previous studies have also utilized mutual information for designing ansatze \cite{qida,zhang2021mutual}; however, the optimal circuit configuration is found using repeated permutations. In QIIA, due to the prior analysis of quantum correlations, mainly entanglement, between the $i^{th}$ qubit and each of the remaining qubits, random gate permutations become unnecessary, resulting in a significant reduction in the complexity of the parametrized circuit. The entire construction becomes fully deterministic, requiring no stochastic sampling. Due to practical computational limitations, we have employed the QIIA with only two blocks in our numerical simulation. 

\begin{table}[tbh]
\caption{\label{tab:energies_for Be like group}
Calculated total energies (in Hartree) for different atomic systems. $E_{DF}$ refers to the DF energy, while $E_{MBPT}^{(2)}$ is the energy obtained using many-body perturbation theory to first-order correction in wave function i.e. the second order correction in energy (referred to also as MP2 in the literature). $E_{CAS-CI}$, $E_{UCCSD}$, and $E_{QIIA}$ are energies calculated using CAS-CI (benchmark value), UCCSD-VQE, and the QIIA-VQE approaches, respectively. $2e^-$,8q refers to a 2-electron, 8-qubit computation. }
\begin{ruledtabular}
\begin{tabular}{lcccccc}
System & $E_{DF}$ & $E_{MBPT}^{(2)}$ & $E_{CAS-CI}$ & $E_{UCCSD}$ & $E_{QIIA}$  \\
\hline
 &  & & 2$e^{-}$, 8q & &  \\ 
 \hline
B$^{+}$ &  -4.97128 & -4.99141 &  -4.99429 &  -4.99429 &  -4.99429  &  \\
C$^{2+}$ & -7.59141 & -7.622404 & -7.64357 &  -7.64357 & -7.64357 \\
N$^{3+}$  & -10.70616 & -10.74533 & -10.78770  &  -10.78770 &  -10.78770  \\
O$^{4+}$  & -14.31962& -14.36600 & -14.43026 & -14.43026 & -14.43026 \\
\hline
 &  & & 4$e^{-}$, 10q & &  \\ 
 \hline
B$^{+}$ &  -24.24516 & -24.26551&-24.28448  & -24.28448 & -24.28441  \\
C$^{2+}$ & -36.42514 & -36.45668& -36.48698 & -36.48698 & -36.48690  \\
N$^{3+}$  & -51.11448 &-51.15451 &  -51.19277 &  -51.19277 & -51.19271 \\
O$^{4+}$  & -68.31439 & -68.36190& -68.40696 & -68.40696 & -68.40684 \\
\end{tabular}
\end{ruledtabular}
\end{table} 

\begin{figure}[tbh]
  \centering
 \includegraphics[scale=.38]{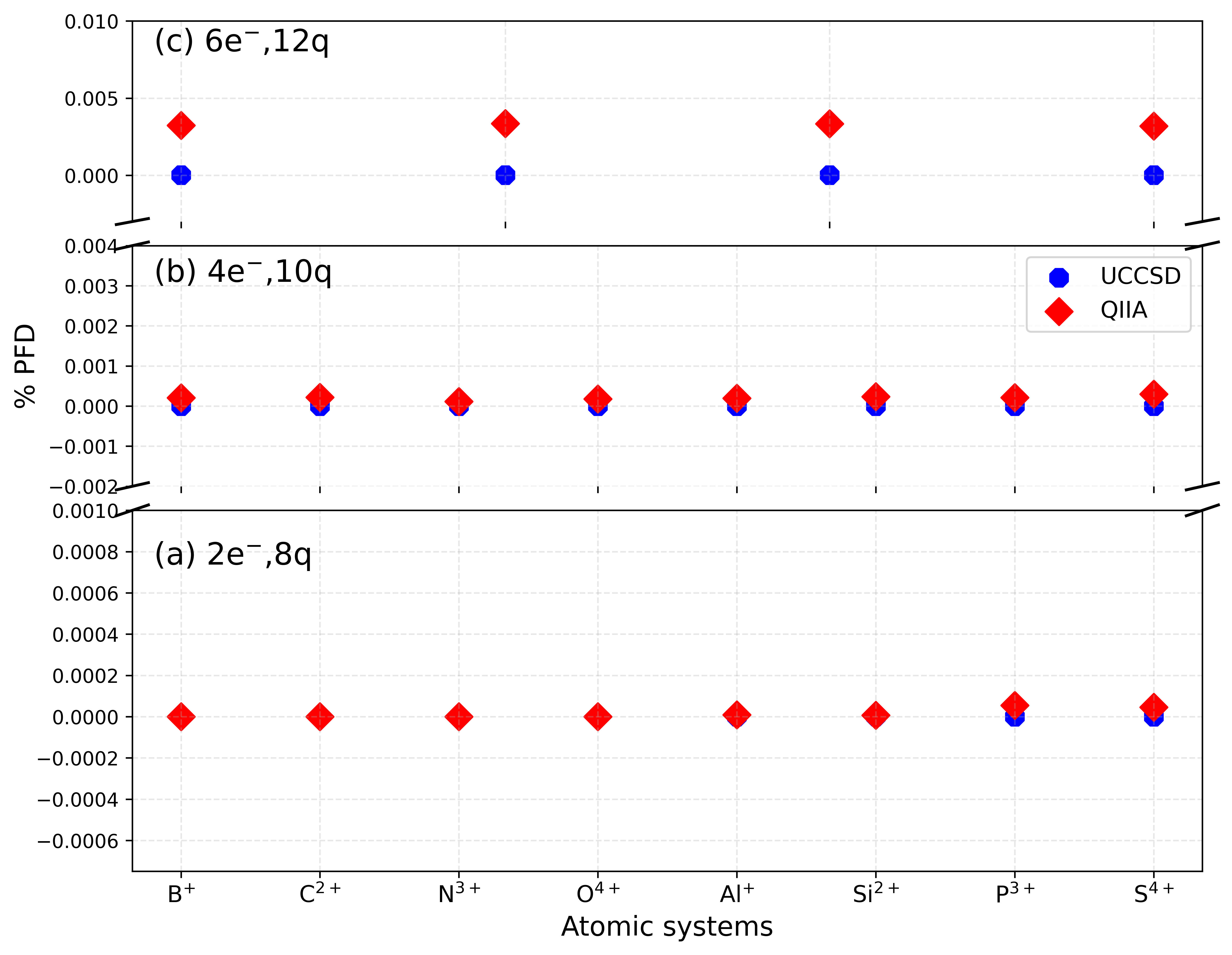}
\caption{Plots showing a comparison between the ground-state energies computed using UCC and QIIA ansatz with two blocks for various atomic systems. PFD refers to the percentage fraction difference from the CAS-CI energy.}
\label{fig: pfd_figure}
\end{figure}

In this work, we have calculated the ground-state energies of four Be-like systems with the chosen active spaces: $2s^2 2p^0$ ($0$ represents the unoccupied orbitals) corresponding to 2 electrons and 8 qubits ($2e^-$,8q), and $1s^2 2s^2 2p^0$ for 4 electrons and 10 qubits ($4e^-$,10q). Similarly, for four Mg-like systems, the active spaces are: $3s^2 3p^0$ for ($2e^-$,8q), $2p_{1/2}^2\, 3s^2 3p^0$ for ($4e^-$,10q), and $2p_{3/2}^4\, 3s^2 3p^0$ for ($6e^-$,12q). 
The input state to the ansatz is set to be the DF state. All our noiseless simulations are carried out using Qiskit $1.4.2$ \cite{javadi2024quantum} 's state-vector backend. For our work, we use the Sequential Least Squares Programming (SLSQP) \cite{kraft1988software} optimizer, with the threshold for convergence set to $10^{-6}$ throughout. The initial parameters in the optimization are randomly chosen. Table~\ref{tab:energies_for Be like group} presents the total ground-state energies (in Hartree) for Be-like ions, computed using various methods across the two active spaces. The results show that the QIIA ansatz, informed by the MBPT first-order wavefunction correction, achieves excellent agreement with CAS-CI reference energies, validating the accuracy of the method and its ability to capture electron correlation effects accurately. Additional simulation results for other atomic systems are provided in Section 4 of the Supplementary Material (see Tables~\ref{tab:8qubitcalcenergies}, \ref{tab:10qcalcenergy}, and \ref{tab:12qcalculationenergies}).
Figure~\ref{fig: pfd_figure} presents the consolidated results for all the atomic systems considered, where the percentage fractional difference (PFD) of ground-state energies computed using QIIA and UCCSD, benchmarked against CAS-CI across different atomic systems and all the active spaces. 
QIIA consistently achieves high accuracy, with deviations under 0.005\%, closely matching the CAS-CI results. It is interesting to see that the QIIA ansatz uses very few resources as compared to the UCCSD ansatz while maintaining good accuracy even up to 12-qubit calculations. The resources taken up in the systems considered for numerical simulations can be found in Figure~\ref {fig:Resources_required} of the Supplementary Material. 
In terms of scaling, the number of parameters scales as \(\sim \mathcal{O}(N^4)\) for UCCSD; however, QIIA requires only \(\sim \mathcal{O}(N^3)\) parameters. Furthermore, the two-qubit gate count reduces from UCCSD's \(\sim \mathcal{O}(N^5)\) to \(\sim \mathcal{O}(N^2)\) for our QIIA ansatz. 

Furthermore, we perform the quantum hardware computations of the two atoms, $\text{P}^{3+}$ and $\text{S}^{4+}$, with an active space of $2p_{1/2}^2\, 3s^2 3p^0$ for four electrons and ten qubits (4e, 10q). For hardware calculation, we first perform the VQE on a classical computer with our QIIA ansatz to obtain the converged amplitudes, which we then use to construct the 
circuit and measure the Hamiltonian on a quantum computer to obtain the ground-state energy. The computations were performed on IBM's Sherbrooke machine having an Eagle R3 processor, which employs a native gate set comprising \{ECR, ID, RZ, SX, X\}. Other details of the hardware can be found in Section 5 of the Supplementary Material.
Obtaining good results on the current noisy hardware requires a workflow with several resource-reduction steps, both in the Hamiltonian and the wavefunction. The number of terms in the Hamiltonian scales as $N^{4}$, where N is the number of spin orbitals, which means that many circuits have to be measured per iteration. This will be difficult to implement on hardware due to error accumulation. 
To reduce the number of terms in the Hamiltonian, we apply a qubit pair-wise commuting grouping called cliques \cite{chawla2025relativistic} that allows simultaneous measurement of commuting Pauli terms. For the 10-qubit Hamiltonian of  $\text{P}^{3+}$ and $\text{S}^{4+}$, the 464 terms in the qubit Hamiltonian are grouped into 209 cliques. Every term is analyzed qubit-wise, and suitable gates are applied to each qubit to rotate the state to the shared eigenbasis of qubit-wise commuting terms \cite{chawla2025vqe}. Thus, only one term from each clique has to be measured in hardware. Further, if two cliques have the same energy contribution, we term them supcliques \cite{chawla2025vqe}, 209 cliques are grouped into 18 supercliques ($\mathcal{S}$) for $\text{P}^{3+}$ and 14 for $\text{S}^{4+}$, out of which we measured only the two dominant supercliques ($\mathcal{S}_{0}$ having clique 0 and $\mathcal{S}_{1}$ having clique 1 and 2) in the hardware for both cases.

Once the ansatz is constructed using the converged parameters from the classical simulation, it undergoes resource reduction strategies to reduce gate counts (especially the number of two-qubit gates) in the circuit. Starting from the entanglers of QIIA ansatz, if the value of $\theta$ in $R_{X}$  and $R_{Z}$ gates is small (at the level of $10^{-3}$), we remove them. Then, the circuit is passed to the Qiskit L3 \cite{javadi2024quantum}$\Rightarrow$ Pytket \cite{sivarajah2020t} $\Rightarrow$ Qiskit L3 routines. After all these steps, the CX gate count reduces from 72 to 48, and the circuit depth reduces from 184 to 80. To execute this reduced QIIA ansatz on IBM hardware, we translate it to the instruction set architecture \cite{javadi2024quantum} circuit, which heavily depends on the quantum computer architecture beneath. To reduce the two-qubit gate on hardware, we use the approximation degree function provided by Qiskit to add another layer to our quantum circuit optimization \cite{javadi2024quantum}. During the computation, we employed error suppression techniques, including dynamical decoupling \cite{viola1999dynamical} to suppress the effects of decoherence on the data qubits and Pauli twirling \cite{bennett1996purification} to suppress the coherent noise. 

The quantum hardware results from the resource-reduced circuits, and where we consider only the supercliques $\mathcal{S}_0$ and $\mathcal{S}_1$ are shown in Table~\ref{tab:superclique_energy} and \ref{tab:energy_comparison_hardware} of Section 4 of the Supplementary Material section. We see that in the noiseless calculations, both the supercliques' energies from the state-vector and QASM results are in close agreement with each other.
$\mathcal{S}_0$ consists of Pauli terms with \textit{I} and \textit{Z} operators, allowing it direct computational basis measurement and particle-number postselection, due to which we obtain an accuracy of \(99.95\%\) for P\(^{3+}\) and \(99.99\%\) for S\(^{4+}\) on hardware with respect to the state-vector energy.
However, not only is the energy contribution from $\mathcal{S}_1$ much smaller, but it also consists of Pauli terms with \textit{X}, \textit{Y}, and \textit{Z} operators, which requires basis rotations for measurement (thus not accommodating our postselection scheme). Therefore we only obtain an accuracy of 13.37\% for P\(^{3+}\) and \(10.85\%\) for S\(^{4+}\) relative to the state-vector energy. This in turn adversely impacts our energy estimates, as seen in Table \ref{tab:energy_comparison_hardware}. We anticipate that in the foreseeable future, quantum hardware advances will alleviate this issue.

In summary, we have introduced a shallow-depth quantum information-inspired ansatz for VQE algorithm. Rather than taking recourse to a heuristic approach to design the parametrized circuits as reported in many previous works \cite{tilly2022variational}, we employ quantum information-theoretic quantities, i.e. von Neumann entropy and quantum mutual information, to build unique blocks and determine the exact positions of the two-qubit entangling gates in the circuit. This results in very accurate ground-state energies of the systems of interest. Our proposed ansatz is therefore more efficient compared to UCC-VQE. The match gate-type particle-conserving entangler keeps the number of particles fixed during the computation, thereby ensuring that the results remain physically correct. QIIA is versatile and can be applied to other quantum many-electron systems. It only relies on a reasonable guess of the target state to arrive at an efficient parametrized circuit using the von Neumann entropy and mutual information data corresponding to the state. 

\textit{Acknowledgment}
AK and PD are grateful to Sudindu Bikash Mandal for useful discussions. AK also wants to acknowledge Tushti Patel and Palak Chawla for their meaningful suggestions. BKS and VSP acknowledge support from the CRG grant
(CRG/2023/002558).
\bibliographystyle{apsrev4-2}
\bibliography{References}


\newpage
\onecolumngrid
\appendix
\section*{Supplementary Material}
\subsection{Many-body perturbation theory corrected wavefunction}


\begin{table}[htbp]
\centering
\caption{Dirac–Fock (DF) reference states and MBPT first-order corrected wavefunctions are obtained for different active spaces of the studied atomic system. For example, \(2s^2 2p^0\) denotes an occupied \(2s\) orbital and virtual \(2p\) orbitals, similar notation is used for other configurations. These normalized MBPT-corrected wavefunctions are used to compute subsystem von Neumann entropies and mutual information, providing insight into electron correlation and orbital entanglement.}

\begin{tabular}{lcl}
\hline\hline
Active Space  & DF state & MBPT first-order corrected wavefunction \\
\hline
(2e,8q) 
    & $2s^2 2p^0$ &
    $\begin{aligned}[t]
    & \phantom{{}+}0.996389\, \ket{00010001} 
    - 0.020173\, \ket{00100010} \\
    & - 0.024871\, \ket{01000100} 
    + 0.078577\, \ket{10001000}
    \end{aligned}$ \\
    
    & $3s^2 3p^0$ &
    $\begin{aligned}[t]
    & \phantom{{}+}0.997060\, \ket{00010001} 
    - 0.023200\, \ket{00100010} \\
    & - 0.023086\, \ket{01000100} 
    + 0.069257\, \ket{10001000}
    \end{aligned}$ \\
\hline
(4e,10q) 
    & $1s^2 2s^2 2p^0$ & 
    $\begin{aligned}[t]
    & \phantom{{}+}0.996590\, \ket{0001100011} 
    - 0.024884\, \ket{0010100101} \\
    & - 0.000266\, \ket{0011000110} 
    - 0.024876\, \ket{0100101001} \\
    & - 0.000266\, \ket{0101001010} 
    + 0.074628\, \ket{1000110001} \\
    & + 0.000799\, \ket{1001010010}
    \end{aligned}$ \\
    
    & $2p_{1/2}^2 3s^2 3p^0$ & 
    $\begin{aligned}[t]
    & \phantom{{}+}0.997060\, \ket{0001100011} 
    - 0.023227\, \ket{0010100101} \\
    & - 0.003415\, \ket{0011000110} 
    - 0.023057\, \ket{0100101001} \\
    & - 0.000137\, \ket{0101001010} 
    + 0.069170\, \ket{1000110001} \\
    & + 0.000411\, \ket{1001010010}
    \end{aligned}$ \\
\hline
(6e,12q) 
    & $2p_{3/2}^4 3s^2 3p^0$ & 
    $\begin{aligned}[t]
    & \phantom{{}+}0.997050\, \ket{000111000111} 
    - 0.023200\, \ket{001011001011} \\
    & + 0.000329\, \ket{001101001101} 
    - 0.000110\, \ket{001110001110} \\
    & - 0.023086\, \ket{010011010011} 
    - 0.002927\, \ket{010110010110} \\
    & + 0.069257\, \ket{100011100011} 
    - 0.002927\, \ket{100101100101}
    \end{aligned}$ \\
\hline\hline
\end{tabular}
\label{tab:mbpt_wavefunctions}
\end{table}

\subsection{Use of entropic measures of quantum correlations}
To construct the hardware-efficient ansatz, we have made use of the quantum correlation structure in the initial multi-qubit state. We have considered two entropic measures of quantum correlations -- von Neumann entropy and mutual information -- to find out how a single qubit is entangled with the other qubits. In quantum information theory, these two measures of quantum correlations are often used in bipartite scenarios. For a single quantum system $A$ in a state $\rho^A$, the von Neumann entropy signifies the amount of inherent randomness and is defined as,
\begin{equation}
    S(\rho^A) = -\mbox{Tr}~\{\rho^A~\mbox{log}~\rho^A\}  
\end{equation}
If $\rho^A$ is the reduced state of a pure entangled state $\rho^{AB}$, then the von Neumann entropy of system $A$ defined above is a measure of entanglement present in the joint state. The state $\rho^{AB}$ is maximally entangled if $S(\rho^{A}) = 1$, whereas, it is separable if $S(\rho^{A}) = 0$. The von Neumann entropy is non-negative for any density operator, i.e., $S(\rho)\geq 0, \forall~\rho$; its minimum value is zero when the density operator is a pure state. 

Another entropic measure of quantum correlations that we have used in our work is the quantum mutual information. In the classical world as well, mutual information is an informational measure of correlations. For a bipartite state $\rho^{AB}$ consisting of subsystems $A$ and $B$, the quantum mutual information is defined as,
\begin{equation}
    I(A:B) = S(\rho^A) + S(\rho^B) - S(\rho^{AB})
\end{equation}
where, $S(\rho^A)$ and $S(\rho^B)$ are the von Neumann entropy of the reduced states $\rho^A$ and $\rho^B$, respectively, and $S(\rho^{AB})$ is the entropy of the joint state. Quantum mutual information is non-negative for any density operator $\rho^{AB}$, and it plays a vital role in measuring the classical and quantum correlations in the quantum world.

In this work, we first use the von Neumann entropy to find out the amount of bipartite entanglement in the state $\rho^{i|remaining}$, where $i$ refers to the $i^{th}$ qubit and 'remaining' refers to all other qubits. Later, we use quantum mutual information to evaluate the amount of pairwise correlations between the $i^{th}$ qubit and all the remaining qubits.     
\begin{table}[t]
\caption{\label{tab:b+top mutual information}
For the B$^{+}$ \((2e^-,8q)\) system, the MBPT first-order corrected wavefunction is used to compute single-qubit von Neumann entropies \((S_i)\) and the full set of mutual information \((I_{i,j})\) values between qubit pairs. The table lists the top mutual information pairs in the whole system and shows which of these are captured within the two entangling blocks of the QIIA ansatz. Results indicate that the two-block structure covers most of the strongest correlations in the system.
As it is an 8-qubit system, a total of 8 entropies are there, which means 8 blocks could be constructed, and we have taken only two blocks corresponding to \(S_0\) and \(S_4\)}. 
\begin{ruledtabular}
\begin{tabular}{lccccc}
System & MBPT first-order  & von Neumann  & Top Mutual  & Two blocks of QIIA &\\
& corrected wavefunction & entropy & information pairs &&\\
\hline
B$^{+}$  & 
$\begin{aligned}[t]
& \phantom{{}+}0.996389\, \ket{00010001} 
- 0.020173\, \ket{00100010} \\
& - 0.024871\, \ket{01000100} 
+ 0.078577\, \ket{10001000}
\end{aligned}$ 
& $\begin{aligned}[t]
&S_0 = 0.05879 \\
&S_4 = 0.05879 \\
&S_3 = 0.04972 \\
&S_7 = 0.04972 \\
&S_1 = 0.00749 \\
&S_5 = 0.00749 \\
&S_2 = 0.00749 \\
&S_6 = 0.00749 \\
\end{aligned}$  &  
$\begin{aligned}[t]
&I_{(0,4)}: 0.05879 \\
&I_{(3,7)} 0.04972 \\
&I_{(0,3)}: 0.04506 \\
&I_{(0,7)}: 0.04506 \\
&I_{(3,4)}: 0.04506 \\
&I_{(4,7)}: 0.04506 \\
&I_{(1,5)}: 0.00749 \\
&I_{(2,6)}: 0.00749 \\
&I_{(0,1)}: 0.00450 \\
&I_{(0,5)}: 0.00450 \\
&I_{(1,4)}: 0.00450 \\
&I_{(4,5)}: 0.00450 \\
&I_{(2,4)}: 0.00450 \\
&I_{(0,2)}: 0.00450 \\
&I_{(0,6)}: 0.00450 \\
&I_{(4,6)}: 0.00450
\end{aligned}$ 
&
$\begin{aligned}[t]
&I_{(0,4)}: 0.05879 \\
&I_{(0,3)}: 0.04506 \\
&I_{(0,7)}: 0.04506 \\
&I_{(3,4)}: 0.04506 \\
&I_{(4,7)}: 0.04506 \\
&I_{(0,1)}: 0.00450 \\
&I_{(0,5)}: 0.00450 \\
&I_{(1,4)}: 0.00450 \\
&I_{(4,5)}: 0.00450 \\
&I_{(2,4)}: 0.00450 \\
&I_{(0,2)}: 0.00450 \\
&I_{(0,6)}: 0.00450 \\
&I_{(4,6)}: 0.00450
\end{aligned}$ &  \\
\end{tabular}
\end{ruledtabular}
\end{table}

\subsection{Match gate entangler}

Let the matrices \( A \) and \( B \) be in \( SU(2) \)
\[
A = \begin{bmatrix}
p & q \\
r & s
\end{bmatrix}, \quad
B = \begin{bmatrix}
w & x \\
y & z
\end{bmatrix}, \tag{6}
\]

with \( \det(A) = \det(B) \). Then, the two-qubit matchgate \( G(A, B) \) is defined as follows:

\[
G(A, B) =
\begin{bmatrix}
p &   &   &  q \\
  & w & x &    \\
  & y & z &    \\
r &   &   & s
\end{bmatrix}. \tag{7}
\]

We are inspired by the work of Bassman \textit{et al.}~\cite{bassman2022constant}, where they presented constant-depth circuits for different Hamiltonian interactions. Since the proposed QIIA ansatz should preserve the particle number in the system, we choose a match gate entangler of the form:

\[
U_{\text{ent}} =
\begin{bmatrix}
\cos\frac{\theta_1 - \theta_2}{2} & 0 & 0 & -i\sin\frac{\theta_1 - \theta_2}{2} \\
0 & \cos\frac{\theta_1 + \theta_2}{2} & -i\sin\frac{\theta_1 + \theta_2}{2} & 0 \\
0 & -i\sin\frac{\theta_1 + \theta_2}{2} & \cos\frac{\theta_1 + \theta_2}{2} & 0 \\
-i\sin\frac{\theta_1 - \theta_2}{2} & 0 & 0 & \cos\frac{\theta_1 - \theta_2}{2}
\end{bmatrix}
\]
and to reduce the number of parameters in the ansatz and make it particle number conserving, we set \( \theta_1 = \theta_2 \).
\[
U_{\text{ent}} =
\begin{bmatrix}
1 & 0 & 0 & 0 \\
0 & \cos{\theta} & -i\sin{\theta} & 0 \\
0 & -i\sin{\theta} & \cos{\theta} & 0 \\
0 & 0 & 0 & 1
\end{bmatrix}
\]

\begin{figure}[tbh]
  \centering
 \includegraphics[scale=.26]{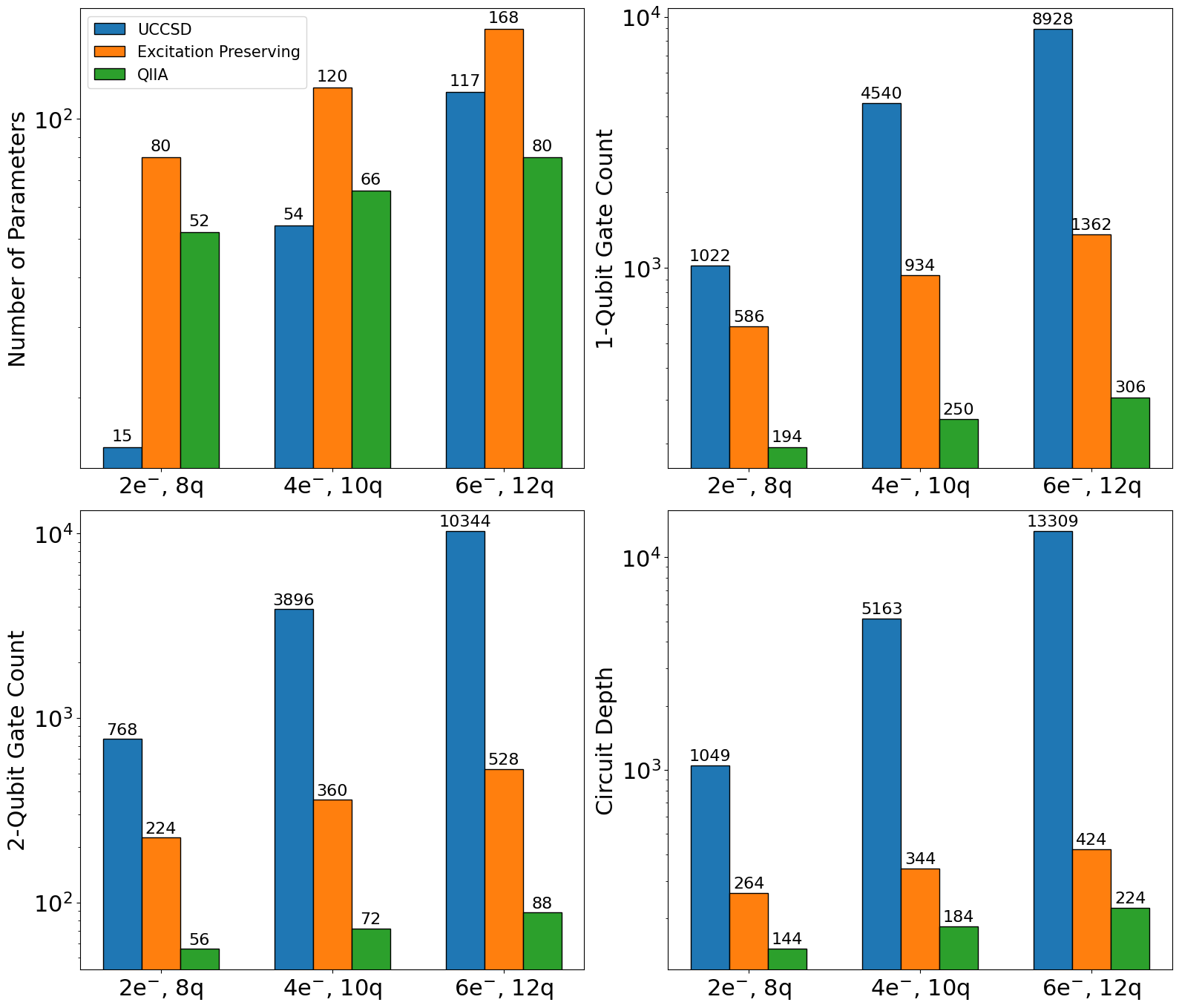}
\caption{Comparison of the quantum resource requirements for UCCSD (blue), excitation-preserving ansatz with full entanglement and two repetition (orange), and QIIA  ansatz with two blocks (green) across the three different active spaces: (2$e^{-}$, 8q), (4$e^{-}$, 10q), and (6$e^{-}$, 12q). 
The panels show (top left) the number of variational parameters, (top right) 1-qubit gate count, (bottom left) 2-qubit gate count, and (bottom right) total circuit depth. 
}
\label{fig:Resources_required}
\end{figure}

By preserving the particle number of the system and taking into account only those qubit pairs that share the maximal amount of quantum correlations between them, the ansatz significantly reduces the quantum resource requirements compared to standard problem-agnostic HEAs. For instance, excitation-preserving HEAs typically consist of layers of $R_z$ rotations interleaved with two-qubit entangling gates generated by $XX+YY$ interactions. Other approaches, such as the Hamiltonian Variational Ansatz\cite{park2024hamiltonian}, Hardware Symmetry-preserving Ansatz,\cite{lyu2023symmetry}, and Entanglement-variational Hardware-efficient Ansatz\cite{wang2024entanglement}, employ variational $XX$, $YY$, and $ZZ$ gates to construct entanglement, but their approach is different from our QIIA.  However, the decomposition of each of the $XX$, $YY$, and $ZZ$ gates requires two CNOT gates, which makes 6 CNOT gates per use of the entangler. 
While using the proposed matchgate-type entangler for $XX$, $YY$ interaction, it only uses 2 CNOT gates per entangler to preserve the particle number in the calculation, therefore reducing the resource significantly compared to other HEA and UCCSD. The heuristic excitation-preserving wave function ansatz is given in Qiskit\cite{javadi2024quantum}, which we are using for comparison in Figure \ref{fig:Resources_required}.

\begin{figure*}[tbh]
  \centering
 \includegraphics[scale=.24]{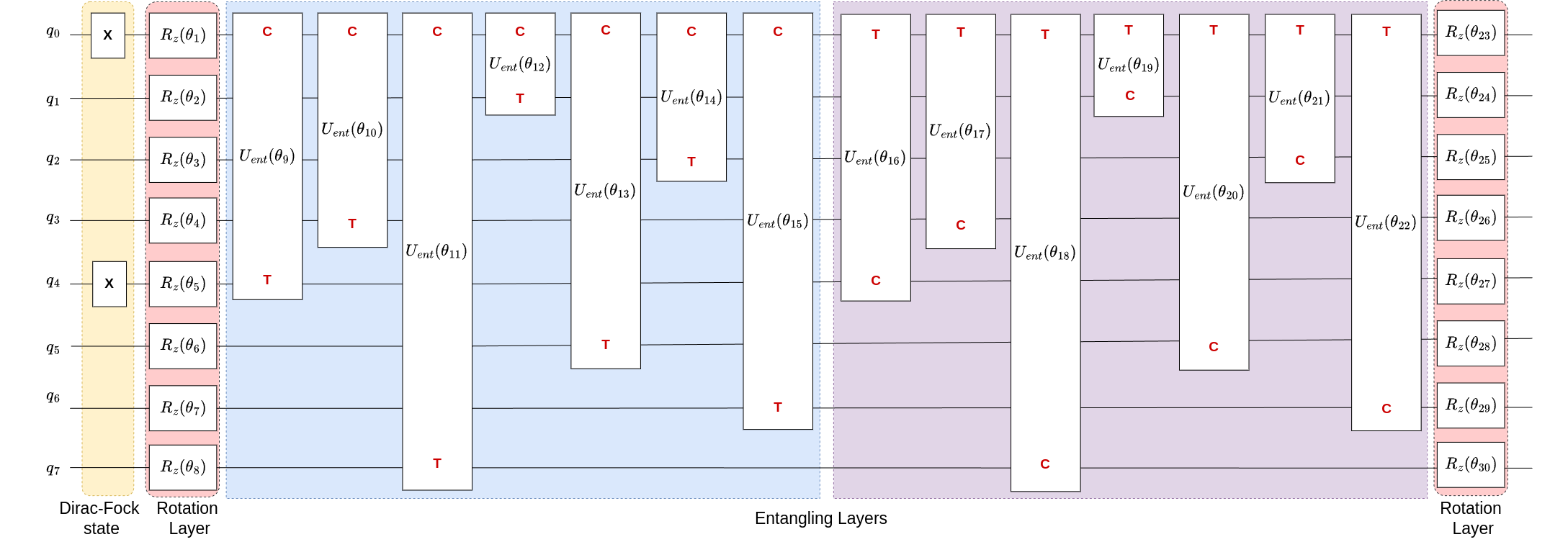}
\caption{A single block of the parametrized circuit corresponding to the proposed ansatz for a 2-electron 8-qubit system. The block is dedicated to $q_{0}$ as it has the highest von Neumann entropy, so the entanglers connect $q_{0}$ and the remaining qubits. The letters `C' and `T' indicate the qubits that are set as control and target, respectively. The circuit for $U_{ent}(\theta_i)$ is given in Figure \ref{fig:flowchart1}. } 
\label{fig:proposed_quantum_circuit}
\end{figure*}

\subsection{Classical simulation}
In the DF calculations, we employ Gaussian-type orbitals (GTOs) as the single-particle basis functions. Specifically, we use a universal basis \cite{dyall1984dirac} set, where the $i^{th}$ function (or basis) is defined as $g_{i}(r) := r^{l_{i}} e^{-\alpha _{i}r^{2}}$, $l_{i}$ being the angular momentum quantum number and $\alpha _{i}$ the exponent of the Gaussian. The distance from the center of the Gaussian to the origin is given by $r$. Further, $\alpha _{i}$ can be expressed as $\alpha _{i} = \alpha_{0}\beta^{i-1}$, for some constants $\alpha_{0}$ and $\beta$. Since we truncate our Hilbert space, we carefully choose the value of $\alpha_{0}$ and $\beta$ that still give the optimal single particle energies as well as the total energy ($\alpha_{0}$ = 0.0009 and $\beta$ = 2.15). The atomic orbitals are expressed within the basis set as a linear combination of basis functions\cite{battaglia2020distributed}. We perform DF calculations using 40 $s_{1/2}$, 39 $p_{1/2}$, 39 $p_{3/2}$, 38 $d_{3/2}$, 38 $d_{5/2}$ basis functions \cite{huzinaga1985basis}. 
After constructing the electronic Hamiltonian, we map it onto qubit operators via the Jordan–Wigner transformation and prepare the QIIA ansatz for VQE simulations using the state-vector backend. Results for Be-like and Mg-like atomic systems are summarized in Tables~\ref{tab:8qubitcalcenergies}, \ref{tab:10qcalcenergy}, and \ref{tab:12qcalculationenergies}. For benchmarking, we compare the VQE energies obtained using just two blocks of QIIA with those from the UCCSD ansatz and classical complete active space configuration interaction (CAS-CI) calculations, which account for all possible excitations within a chosen active space, offering an exact diagonalization in that subspace.CAS-CI offers a full CI treatment confined to this restricted space.

\begin{table}[h]
\caption{\label{tab:8qubitcalcenergies}
Calculated total ground-state energies (in Hartree) for selected atomic systems in the \((2e^{-}, 8q)\) active space using various methods. The QIIA ansatz used in the VQE calculations includes two entangling blocks. 
Percentage fractional differences (PFD) are computed with respect to the CAS-CI results, which serve as the reference for how much energy the method loses to CAS-CI.}
\begin{ruledtabular}
\begin{tabular}{lcccccccc}
System & $E_{DF}$ & $E_{CAS-CI}$ & $E_{MBPT}^{(2)}$ & PFD ($E_{MBPT}^{(2)}$) & $E_{UCCSD}$ & PFD ($E_{UCCSD}$) & $E_{QIIA}$ & PFD ($E_{QIIA}$) \\
\hline
B$^{+}$ &  -4.97128 &  -4.99429 & -4.99141 & 0.05767 & -4.99429 & 0.00000 &  -4.99429  & 0.00000 \\
C$^{2+}$ & -7.59141 &  -7.64357 & -7.62240 & 0.27691 & -7.64357 & 0.00000 & -7.64357 &  0.00000\\
N$^{3+}$  & -10.70616 &  -10.78770  & -10.74533 & 0.39276 & -10.78770 & 0.00000 & -10.78770 & 0.00000 \\
O$^{4+}$  & -14.31962 & -14.43026 & -14.36600 & 0.44531 &-14.43026 & 0.00000 &-14.43026 & 0.00000 \\
Al$^{+}$  & -11.51436 &  -11.51748 & -11.52566  & -0.07102 & -11.51748 & 0.00000 &-11.51747 &  0.00009 \\
Si$^{2+}$  & -14.78301 & -14.79066 & -14.80088 & -0.06910 &-14.79066 & 0.00000 &-14.79065 &  0.00007\\
P$^{3+}$  & -18.17258 &  -18.18542 & -18.19513 & -0.05339  &-18.18542 &0.00000 &-18.18541 &  0.00005 \\
S$^{4+}$ & -21.72867   & -21.74731 & -21.75507 & -0.03572 &-21.74731 & 0.00000 &-21.74730 & 0.00005 \\
\end{tabular}
\end{ruledtabular}
\end{table}
\newpage
\begin{table}[t]
\caption{\label{tab:10qcalcenergy}
Calculated total ground-state energies (in Hartree) for selected atomic systems in the \((4e^{-}, 10q)\) active space using various methods. The QIIA ansatz used in the VQE calculations includes two entangling blocks. 
Percentage fractional differences (PFD) are computed with respect to the CAS-CI results, which serve as the reference for how much energy the method loses to CAS-CI.}
\begin{ruledtabular}
\begin{tabular}{lcccccccc}
System & $E_{DF}$ & $E_{CAS-CI}$ & $E_{MBPT}^{(2)}$ & PFD ($E_{MBPT}^{(2)}$) & $E_{UCCSD}$ & PFD ($E_{UCCSD}$) & $E_{QIIA}$ & PFD ($E_{QIIA}$) \\
\hline
B$^{+}$ &  -24.24516 & -24.28448 & -24.26551 & 0.07812  & -24.28448 & 0.00000 & -24.28441 & 0.00029 \\
C$^{2+}$ & -36.42514 & -36.48698 & -36.45668 &  0.08304 &  -36.48698 & 0.00000 & -36.48690 & 0.00022 \\ 
N$^{3+}$  & -51.11448 &  -51.19277 & -51.15451 & 0.07474 &-51.19277  & 0.00000 & -51.19271 & 0.00012 \\
O$^{4+}$  & -68.31439 & -68.40696 &-68.36190 & 0.06587 & -68.40696 &0.00000 & -68.40684 & 0.00018  \\
Al$^{+}$  &  -45.75368 & -45.90465 &-45.76500 & 0.30422 & -45.90465 & 0.00000& -45.90451 & 0.00030  \\
Si$^{2+}$ & -55.23685 & -55.52703 & -55.25480 & 0.49027 & -55.52703  &0.00000 & -55.52690 & 0.00023 \\
P$^{3+}$  & -65.36936 & -65.78414 & -65.39204  &  0.59604  & -65.78414 &0.00000 & -65.78400 & 0.00021   \\
S$^{4+}$ &  -76.18899 & -76.71452 & -76.21558 & 0.65039 & -76.71452  &0.00000 & -76.71409 & 0.00056  \\

\end{tabular}
\end{ruledtabular}
\end{table}

\begin{table}[t]
\caption{\label{tab:12qcalculationenergies}
Calculated total ground-state energies (in Hartree) for selected atomic systems in the \((6e^{-}, 12q)\) active space using various methods. The QIIA ansatz used in the VQE calculations includes two entangling blocks. 
Percentage fractional differences (PFD) are computed with respect to the CAS-CI results, which serve as the reference for how much energy the method loses to CAS-CI.}
\begin{ruledtabular}
\begin{tabular}{lcccccccc}
System & $E_{DF}$ & $E_{CAS-CI}$ & $E_{MBPT}^{(2)}$ & PFD ($E_{MBPT}^{(2)}$) & $E_{UCCSD}$ & PFD ($E_{UCCSD}$) & $E_{QIIA}$ & PFD ($E_{QIIA}$) \\
\hline
Al$^{+}$  & -73.95504 &  -74.18365 & -73.96647 & 0.29276  & -74.18364 & 0.00001 & -74.18125  & 0.00324  \\
Si$^{2+}$  &  -88.88696  &  -89.31982 & -88.90522 & 0.46417 &-89.31981 & 0.00001 & -89.31682  &  0.00336\\
P$^{3+}$& -104.98313 &   -105.59360 & -105.00634 & 0.55615 &-105.59359 & 0.000009 & -105.59008  & 0.00333 \\
S$^{4+}$ &  -122.27074 &   -123.03490 & -122.29806 & 0.59889 &-123.03489 & 0.000008 & -123.03097  & 0.00319 \\
\end{tabular}
\end{ruledtabular}
\end{table}


Fig.~\ref{fig:Resources_required} presents a comparison of the resource requirements for the proposed QIIA ansatz against the UCCSD ansatz and the excitation-preserving HEA. The comparison is motivated by the fact that both UCCSD and excitation-preserving ansatz conserve particle number, UCCSD by construction, and the excitation-preserving ansatz through the use of $XX + YY$ interactions. QIIA similarly uses $XX + YY$ entanglers but introduces a matchgate-inspired strategy informed by quantum information metrics, offering a more targeted and efficient approach.
To provide a fair comparison, we consider the excitation-preserving ansatz with two repetitions and full entanglement connectivity, as well as the QIIA ansatz with two von Neumann blocks. QIIA achieves the same particle-number conservation while requiring significantly fewer variational parameters, single- and two-qubit gates, and overall circuit depth. This demonstrates that QIIA is both resource-efficient and scalable, making it highly suitable for implementation on near-term quantum hardware.

\subsection{Hardware calculation}

Hardware simulations were performed on IBM's Sherbrooke quantum device equipped with the Eagle R3 processor, using its native gate set $\{$ECR, ID, RZ, SX, X$\}$. The system exhibited median error rates of $6.184 \times 10^{-3}$ (ECR), $2.305 \times 10^{-4}$ (SX), and $2.051 \times 10^{-2}$ (readout), with coherence times of $T_1 = 267.74\,\mu\mathrm{s}$ and $T_2 = 208.12\,\mu\mathrm{s}$. The qubit layout follows a heavy-hex topology with a CLOPS (Circuit layer operations per second) score of 150K. CLOPS is a metric that represents the number of quantum volume circuits a quantum processing unit can execute within a given period.
For hardware evaluation, we adopt a hybrid strategy. First, we perform a full VQE calculation using the QIIA ansatz classically to obtain converged parameters. These parameters are then embedded into a circuit, which is measured on hardware. This way of implementation is different from a full VQE calculation, which involves each iteration on a quantum computer. But because of the high cost and limited quantum time available in hardware, we do not opt for this procedure. 

\begin{table*}[htb]
\caption{For the P$^{3+}$ and S$^{4+}$ systems with active space of \(4e^-, 10q\) the Hamiltonian is decomposed into supercliques \((\mathcal{S})\), which are further divided into cliques containing specific Pauli terms. For each superclique, we list the expectation values from state-vector simulation, QASM simulation, and IBM quantum hardware runs. The total energy is calculated from the contributions of the two supercliques.}

\centering
\begin{ruledtabular}
\begin{tabular}{lcccccr}
 &  & P$^{3+}$  & & & \\ 
 \hline
Superclique ($\mathcal{S}$) & Cliques in $\mathcal{S}_i$ & Terms &$E_{State-vector}$ &$E_{QASM}$ & $E_{Hardware}$ \\ 
\hline
$\mathcal{S}_0$ & Clique 0 & IIIIIIIIII, IIIIIIZIII, $\cdots$, ZZIIIIIIII  & $-64.97721$& $-64.97719$ &  $-64.94349$\\
\hline
$\mathcal{S}_1$ & Clique 1 & IIYZYIIIII, IIIIIIIYZY, $\cdots$, IIIIIZYYII & $-0.41085$& $-0.41084$ &$-0.05491$ \\
      & Clique 2 & IIXZXIIIII, IIIIIIIXZX, $\cdots$, IIIIIZXXII &$-0.41085$ & $-0.41084$ & $-0.05491$ \\
\hline
 & & Total Energy =  & $-65.79891$   & $-65.79887$&
 $-64.99840$ \\
\hline
 &  & S$^{4+}$  & &  &\\ 
\hline
$\mathcal{S}_0$ & Clique 0 & IIIIIIIIII, IIIIIIZIII, $\cdots$, ZZIIIIIIII & $-75.69358$& $-75.69318$ & 
$-75.69271$ \\
\hline
$\mathcal{S}_1$ & Clique 1 & IIYZYIIIII, IIIIIIIYZY, $\cdots$, IIIIIZYYII & $-0.51610$& $-0.51602$ &  $-0.05601$ \\
      & Clique 2 & IIXZXIIIII, IIIIIIIXZX, $\cdots$, IIIIIZXXII & $-0.51610$ & $-0.51602$& $-0.05601$ \\
\hline
 & & Total Energy =  & 
 $-76.72578$ & $-76.72522$ & $-75.80473$ \\
\end{tabular}
\end{ruledtabular}
\label{tab:superclique_energy}
\end{table*}

\begin{table*}[tbh]
\centering
\caption{%
Computed ground-state energies (\( E \)) and percentage fractional differences (PFD) with respect to CAS-CI for P$^{3+}$ and S$^{4+}$ atomic systems using methods such as, Dirac-Fock (\( E_{\text{DF}} \)), MBPT first order wavefunction correction and second order energy correction (\( E_{\text{MBPT}^{(1)}} \)), CAS-CI (\( E_{\text{CAS-CI}} \)), UCCSD (\( E_{\text{UCCSD}} \)), and QIIA with two blocks(\( E_{\text{QIIA}} \)).  
\( E_{\text{QIIA (Reduced)}} \) denotes the reduced QIIA circuit after applying resource reduction on it and 
\( E_{\text{QIIA (Reduced)}^{*}} \) denotes the hardware results from IBM Sheerbroke with a reduced circuit using only two dominant supercliques. PFD calculation is done with respect to the CAS-CI energy.
}

\begin{tabular}{lccc}
\hline\hline
System & Method & Energy \( E \) (a.u.) & \ PFD \\
\hline
\textbf{P$^{3+}$ (4e,10q)} 
    & \( E_{\text{DF}} \) &  -65.36936 & 0.63051 \\
    & \( E_{\text{MBPT$^{(2)}$}} \) & -65.39204 &  0.59604 \\
    & \( E_{\text{CAS-CI}} \) & -65.78414  & --\\
    & \( E_{\text{UCCSD}} \) & -65.78414  & 0.00000\\
    & \( E_{\text{QIIA}} \) & -65.78400  & 0.00021\\
    & \( E_{\text{QIIA (Reduced)}} \) & -65.77433 & 0.01491  \\
    & \( E_{\text{QIIA (Reduced)}^{*}} \) & -64.99840 & 1.19442\\
\hline
\textbf{S$^{4+}$ (4e,10q)} 
    & \( E_{\text{DF}} \) & -76.18899 & 0.68504 \\
    & \( E_{\text{MBPT$^{(2)}$}} \) & -76.21558 & 0.65039 \\
    & \( E_{\text{CAS-CI}} \) & -76.71452 & -- \\
    & \( E_{\text{UCCSD}} \) & -76.71452 &  0.00000\\
    & \( E_{\text{QIIA}} \) & -76.71409 & 0.00056 \\
    & \( E_{\text{QIIA (Reduced)}} \) & -76.70433 & 0.01328 \\
    & \( E_{\text{QIIA (Reduced)}^{*}} \) & -75.80473 & 1.18594 \\
\hline\hline
\end{tabular}
\label{tab:energy_comparison_hardware}
\end{table*}

The experiments were performed for P$^{3+}$ and S$^{4+}$ having configuration of  4-electron, 10-qubits. The resource reduction strategy includes compressing the Hamiltonian into cliques and supercliques ($\mathcal{S}$). For example, P$^{3+}$, the full Hamiltonian had 464 Pauli terms, grouped into 209 cliques and further into 18 supercliques. Only two dominant supercliques were selected for hardware measurements, reducing the number of distinct measurement settings to two.
Further circuit resource reduction is applied to the QIIA ansatz. Small-angle single-qubit gates (e.g., $R_X$, $R_Z$ with $\theta \sim 10^{-3}$) and the corresponding CNOTs were pruned. The circuit was transpiled via Qiskit L3 $\rightarrow$ Pytket $\rightarrow$ Qiskit L3 flow, which reduced the CX gate counts from 72 to 48 and the circuit depth from 184 to 80. The circuit was then compiled to the device’s native gate set, accounting for the heavy-hex connectivity of the hardware. This mapping introduced additional SWAP operations, increasing the two-qubit ECR gate count to 114 and the depth to 770. Applying Qiskit’s approximation passes subsequently reduced the ECR gate count to 17, while maintaining the overall depth at 184. For handling the hardware noise,
Error suppression techniques such as dynamical decoupling (XX-type) and Pauli twirling were used. Measurements from supercliques $\mathcal{S}_0$ and $\mathcal{S}_1$ are shown in Table~\ref{tab:superclique_energy}. 
Table~\ref{tab:energy_comparison_hardware} compares total energy (from two dominant supercliques) with DF, MBPT$^{(2)}$, CAS-CI, UCCSD, and QIIA energies.

\end{document}